# Control of nanomechanical resonances by an electron beam

Toji Thomas[1], Kevin F. MacDonald[1] and Eric Plum[1]

1. Optoelectronics Research Centre, University of Southampton, Highfield, Southampton, SO17 1BJ, UK

**Abstract**: The sensitivity of mechanical resonators to physical quantities such as acceleration, pressure, mass and temperature enables them to underpin sensing and metrology applications. Here, we observe that the resonance frequency of a nanomechanical resonator depends strongly on charging. We show that repulsion between an electron beam and charge accumulated on a nanomechanical cantilever yields a stiffening that increases its resonance frequency, providing a mechanism for controlling resonators and sensing charge. For a cantilever of microscale length and nanoscale cross-section interacting with the electron beam of a scanning electron microscope, we observe a resonance shift on the order of 1% per nanocoulomb.

**Introduction**

Micro- and nanomechanical systems underpin a wide range of applications from biology to fundamental physics [1-4]. The sensitivity of the mechanical response of suitably designed mechanical systems to optical forces, mass, pressure, temperature, charge, electric and magnetic fields, etc., can be applied in ultrasensitive sensors [5-7], tuneable filters [8], the realization of time crystals [9, 10] and the study of fundamental phenomena such as thermal motion [11, 12], Casimir effect, etc. [13]. Techniques developed to detect and image such fast, small-scale motion with high motion sensitivity and spatial and temporal resolution include optical interferometric techniques [14], laser Doppler vibrometry [15], as well as certain super-resolution optical techniques [16], electronic detection techniques [17] and scanning probe methods [12, 18]. Electron-beam probe techniques avoid common limitations of other methods, which are typically either frame rate limited, requiring complex electronic circuitry, or diffraction limited. Scanning electron microscopy (SEM) [12, 19] and transmission electron microscopy (TEM) [20] can give sub-nanometre displacement sensitivity combined with nanoscale spatial and microsecond temporal resolution. However, electron beam–sample interaction introduces challenges, such as electron beam-induced carbon contamination [21, 22] and charging [23, 24]. Charging is the accumulation of incident electrons within the sample and occurs particularly in poorly grounded or non-conductive samples [24]. It introduces imaging artifacts [25, 26], distorts the beam trajectory [27], and, most importantly for this study, perturbs the mechanical behavior of the resonator itself. The mechanical behavior of charged micro- and nanomechanical resonators has been studied in the context of AFM-based techniques [28-32], where the accumulated charge on the surface of the sample can modify its local electrostatic environment, causing a resonance frequency shift of the probing cantilever due to electrostatic repulsion or attraction of the conductive tip. Such capacitive coupling has been used [33-38] to adjust resonance frequencies by modulating an applied voltage, exploiting the associated electrostatic spring softening or stiffening effect [39, 40]. However, in focused electron beam techniques, such charging effects are usually avoided using conductive coatings and grounding schemes [41].

While steps are normally taken in electron microscopy to eliminate or at least minimize charging, here, we explore how it can be exploited. We systematically investigate how electron beam–induced charging affects the resonant behavior of nanoscale mechanical resonators (Figure 1) and explore its potential for tuning of mechanical resonances and detection of charged particle beams. We study the mechanical response of an isolated externally driven cantilever using secondary electron nanomotion metrology [12] for different electron beam positions and currents. We observe an electron-beam-induced blue-shift of the resonator's mechanical resonance which we attribute to an electrostatic stiffening effect arising from the Coulomb interaction between the electron beam and charge deposited in the cantilever. We derive a simple analytical model that describes the dependence of the resonant characteristics of a charged mechanical oscillator on a nearby charged particle beam.

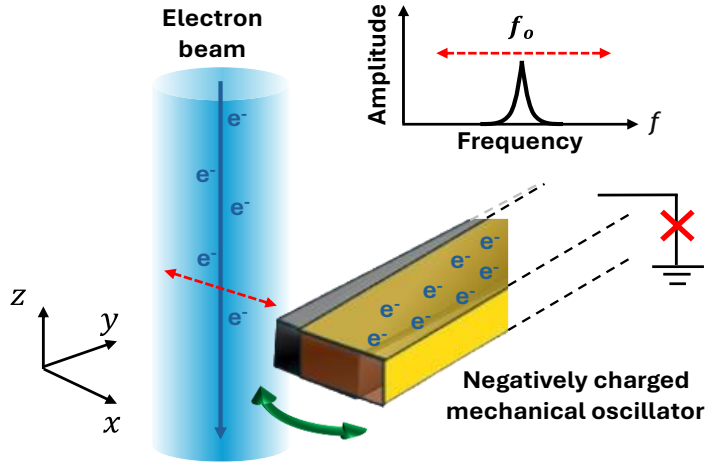

*Figure 1. An isolated mechanical resonator will be charged by a nearby electron beam, causing a Coulomb force between the electron beam and the resonator, which stiffens the resonator. Moving the electron beam closer to (away from) the resonator will blue-shift (red-shift) its mechanical resonances.*

**Experiment**

We study the motion of cantilevers cut from a silicon nitride membrane coated with gold and gallium nitride layers, with and without patterning around their base for electrical isolation (Fig. 2a, see Methods for details). The sample was mounted on a piezo actuator that is used to drive in-plane oscillations of the cantilevers using a sinusoidal driving voltage of 500 mV$_{pp}$. The SEM is then operated in 'spot mode,' where the electron beam with 1.5 nA current and 10 kV acceleration voltage is positioned (blue dots in Fig. 2a) at different distances from the edge of the structure. Cantilever motion with respect to the electron beam modulates the rate of electron impact on the cantilever and thus also the resulting rate of secondary electron generation and the output signal of the electron microscope's secondary electron detector. For a fixed electron beam position and small amplitudes of nanomechanical oscillation, the secondary electron signal modulation is proportional to the oscillation amplitude and measured in our experiments by lock-in detection at the driving frequency. Reference measurements are taken with the electron beam positioned $d=$ 300 nm away from the cantilever edge, a sufficient distance for a negligible influence of the electron beam on the cantilever. We observe the fundamental in-plane resonance frequencies at 332.68 kHz for the isolated cantilever (Fig. 2b) and 407.69 kHz for the grounded cantilever (Fig. 2c), with similar resonance quality factors of 600 and 630, respectively. Resonance frequencies and quality factors were determined by fitting Eq. (10) to the experimental data. Measurements are taken at electron beam distances $d$ from 200 nm to 20 nm from the cantilever edge in 20 nm steps. As the electron beam approaches the edge of the isolated cantilever, its resonance frequency blue-shifts significantly and continuously (Fig. 2b). In case of the grounded cantilever, a blue-shift is also detectable, but it is much smaller and seen only at smaller distances $d$ (Fig. 2c). For example, the blue-shift arising from positioning the electron beam 100 nm from the cantilever edge is 490 Hz for the isolated cantilever and only 30 Hz for the grounded cantilever, a difference of more than an order of magnitude. For closer proximity of the electron beam to the cantilever edge the blue-shift continues to increase for both cantilevers, reaching 1170 and 270 Hz at $d=$ 20 nm, respectively. The much larger frequency shift that occurs for the isolated cantilever compared to the grounded one suggests that the blue-shift may be caused by charging.

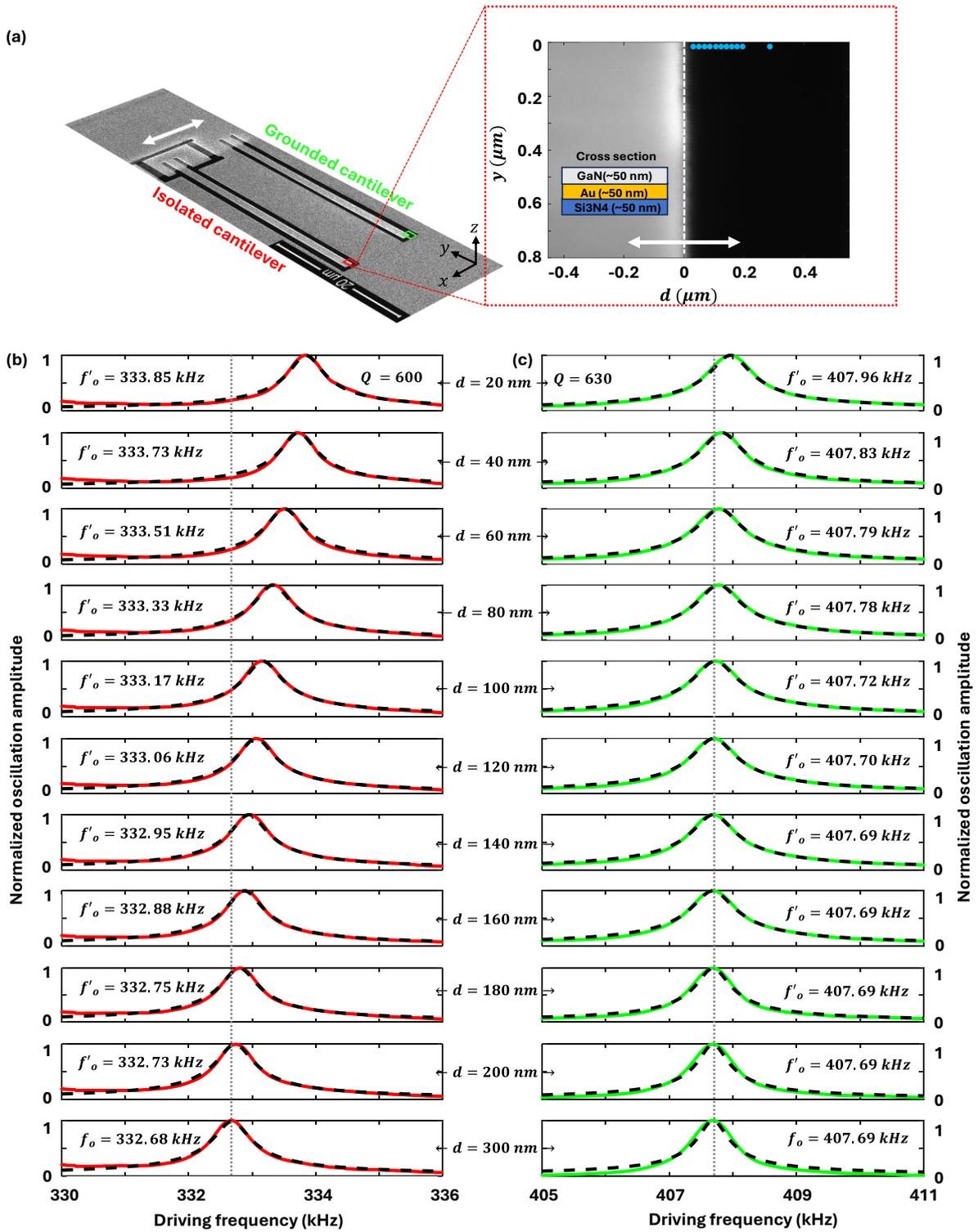

Figure 2: Mechanical resonances detected at different electron beam injection points for isolated and grounded cantilevers. (a) SEM image of the cantilevers, which are nominally identical with the same layers (inset) and dimensions of 32 μm length, 800 nm width, and 150 nm thickness. The conductive gold layer at the base of one cantilever has been milled away to isolate it electrically (red), while the other remains grounded (green). The motion of the cantilevers is probed using the electron beam positioned (blue dots) at different distances $d$ from the edge (white line) of each cantilever near its tip, as indicated on the enlarged SEM image. Lateral oscillations (white double-headed arrow) of the cantilevers are driven by a piezo actuator. (b & c) Oscillation amplitude as a function of driving frequency for the isolated (red) and grounded (green) cantilevers with fits (dashed) according to Eq. (10). As the electron beam injection position approaches the cantilevers, a blue-shift of their resonance frequencies ($f_o$) is observed, which is much larger for the isolated cantilever.

**Analytical model**

Experimental evidence (Fig. 2b, c) suggests that the resonance frequency shift is due to charge build-up on the cantilever and its interaction with the electron beam. The observed blue-shift of the in-plane resonance implies an increase of the cantilever's effective spring constant, i.e. an additional force arising from the charge-electron beam interaction and acting against in-plane ($x$) displacement of the cantilever. The interaction may be understood by considering the force acting on a point charge (the charged cantilever) placed in the vicinity of a line of moving charges (the electron beam). Being essentially a line charge and line current (along $z$), the electron beam generates a radial electric and azimuthal magnetic field [42]. Any contributions from the magnetic Lorentz force due to charge motion (due to in-plane cantilever oscillations or charge flow along the cantilever) can be excluded as the interaction of such charge motion in the $xy$-plane with magnetic field in the $xy$-plane can only yield a Lorentz force along $z$. However, the charged cantilever within the radial electric field of the electron beam will experience a Coulomb force along the direction of its in-plane oscillations ($x$). Indeed, stronger/weaker repulsion for cantilever displacement towards/away from the electron beam may be expected to yield a stiffening effect and thus a blue-shift of the cantilever's in-plane resonance. Here, we will evaluate how the mechanical resonance is affected by the electric force $\vec{F_e}$ on a point charge $q$ (representing the charge on the cantilever) located in the electron beam's radial electric field $\vec{E}$.

Approximating the electron beam as a cylindrical wire of infinite length and infinitesimal radius carrying a positive current $I_e$, Gauss's Law gives the electric field outside the cylindrical charged wire as $\oint \vec{E} \cdot d\vec{A} = Q_{enc}/\varepsilon_0$. For a cylindrical surface of radius $R$ and length $L$ around the electron beam, the enclosed charge is $Q_{enc} = (dQ_{enc}/dL)L = (-I_e/v_e)L$, where $v_e$ is the electron velocity that is determined by the acceleration voltage $V_e$ and given by $v_e = \sqrt{(2eV_e/m_e)}$ in the non-relativistic regime that we consider here ($V_e = 10$ kV). $dQ_{enc}/dL$ is the charge per unit length, $\varepsilon_o$ is the permittivity of free space, $e$ charge of an electron, and $m_e$ mass of an electron. Since the field is cylindrically symmetric, $\vec{E}(2\pi RL) = -I_e L/(v_e \varepsilon_0)\hat{e}_r$, rearranging for $\vec{E}$,

$$\vec{E} = \frac{-I_e}{2\pi R v_e \varepsilon_0} \hat{e}_r \quad (1)$$

The electric force on a charge $q$ in the radial electric field is

$$\vec{F_e} = q\vec{E} = \frac{-qI_e}{2\pi R v_e \varepsilon_0} \hat{e}_r \quad (2)$$

Considering the negative charge $q$ of the cantilever, the electric force between electron beam and cantilever is repulsive. To determine the effect of this repulsive force on the cantilever's fundamental resonance of in-plane oscillation, we take it into account in its equation of motion. We treat the cantilever as a damped harmonic oscillator with mass $m_{eff}$ oscillating about $x_o$ in response to an oscillating driving force and the electric force acting on its charge $q$ within the electric field of the electron beam at a distance $R(t) = R_0 + \Delta x(t)$ (Fig. 3). Where $R_0$ is the equilibrium distance between the charge and the electron beam and $\Delta x(t)$ is the displacement of the cantilever from its equilibrium position $x_o$. The cantilever's position is $x(t) = x_o + \Delta x(t)$ and without electric force $x_o$ shall be 0.

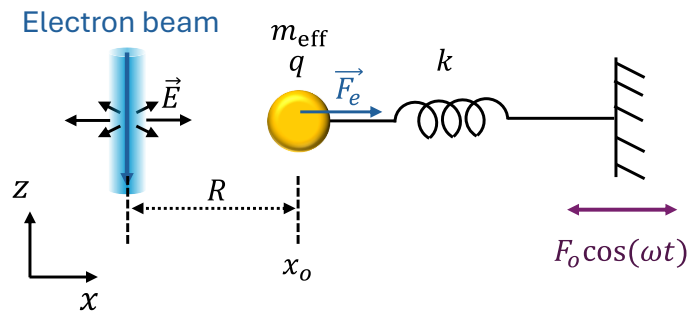

*Figure 3: A charged and driven mechanical oscillator in the presence of an electron beam can be approximated as a charged mass on a spring under forced vibration within the electric field of an infinitesimally thin cylindrical wire carrying a line charge. The Coulomb electrostatic force changes the effective spring constant, resulting in a change of the mechanical resonance frequency.*

The equation of motion of the driven and damped, charged oscillator experiencing a Coulomb force ($F_e$) is

$$m_{\text{eff}}\ddot{x} + b\dot{x} + kx = F_e + F_o \cos(\omega t), \tag{3}$$

where $k$ is the spring constant, $m_{\text{eff}}$ is the effective mass, $b$ is the damping coefficient, $F_o$ and $\omega$ are amplitude and angular frequency of the driving force, and $t$ is the time. Substituting Eq. (2) in (3),

$$m_{\text{eff}}\ddot{x} + b\dot{x} + kx = \frac{-qI_e}{2\pi\varepsilon_o v_e}\left(\frac{1}{R_o + \Delta x}\right) + F_o \cos(\omega t) \tag{4}$$

If $|\Delta x| \ll R_o$, the electrostatic term can be approximated using Taylor expansion $\frac{1}{R_o+\Delta x} = \frac{1}{R_o} - \frac{\Delta x}{R_o^2} + \frac{\Delta x^2}{R_o^3} - \cdots$. Ignoring higher orders,

$$m_{\text{eff}}\ddot{x} + b\dot{x} + kx \approx \frac{-qI_e}{2\pi\varepsilon_o v_e R_o} + \frac{qI_e}{2\pi\varepsilon_o v_e R_o^2}\Delta x + F_o \cos(\omega t) \tag{5}$$

Here, $F_{eo} = \frac{-qI_e}{2\pi\varepsilon_o v_e R_o}$ is the electrostatic force acting on the oscillator at its (new) equilibrium position, i.e. it shifts the equilibrium position from 0 to $x_o \approx F_{eo}/k$. Considering this, substituting $x(t) = x_o + \Delta x(t)$ and rearranging, Eq. (5) simplifies to the equation of motion for the displacement $\Delta x$ relative to the new equilibrium position

$$\ddot{\Delta x} + \frac{b}{m_{\text{eff}}}\dot{\Delta x} + \frac{\left(k - \frac{qI_e}{2\pi\varepsilon_o v_e R_o^2}\right)}{m_{\text{eff}}}\Delta x = \frac{F_o}{m_{\text{eff}}}\cos(\omega t) \tag{6}$$

Eq. (6) has the same form as the equation of motion of a forced oscillator [43],

$$\ddot{x} + \gamma\dot{x} + \omega_o^2 x = \frac{F_o}{m_{\text{eff}}}\cos(\omega t) \tag{7}$$

Comparing coefficients of Eq. (6) and (7) gives the modified natural frequency,

$$\omega_o' = 2\pi f_o' = \sqrt{\frac{\left(k - \frac{qI_e}{2\pi\varepsilon_o v_e R_o^2}\right)}{m_{\text{eff}}}} \approx \omega_o\left(1 - \frac{qI_e}{4\pi\varepsilon_o v_e R_o^2 k}\right) \tag{8}$$

(where the approximation holds for small relative frequency shifts), modified spring constant,

$$k' = k - \frac{qI_e}{2\pi\varepsilon_o v_e R_o^2} \tag{9}$$

and amplitude as a function of driving frequency,

$$A(\omega) = \frac{F_o/m_{\text{eff}}}{\sqrt{\left[\omega_o'^2 - \omega^2\right]^2 + \left(\frac{\omega_o'\omega}{Q}\right)^2}} \tag{10}$$

Eq. (8) and (9) imply that the electrostatic force stiffens the spring as the distance between the negative charge and the electron beam decreases, thus increasing the resonance frequency. Eq. (10) was used to fit the experimental data.

**Evaluating the charge on a cantilever**

The analytical expression for the modified resonance frequency, Eq. (8), can be used to estimate the charge $q$ on the cantilevers in our experiments (Fig. 2 b, c)

$$q = \frac{-2\pi\varepsilon_o v_e R_o^2 k}{I_e}\left(\left(\frac{f_o'}{f_o}\right)^2 - 1\right), \tag{11}$$

where, for small relative frequency shifts $\left|\frac{f'_o}{f_o} - 1\right| \ll 1$ as observed here, $\left(\frac{f'_o}{f_o}\right)^2 - 1 \approx 2\left(\frac{f'_o}{f_o} - 1\right)$, implying that the accumulated charge is proportional to the relative frequency shift. Fig. 4a shows the relative frequency shift as a function of the distance between electron beam and cantilever edge for the isolated (red) and grounded (green) cantilevers, illustrating the much larger electron-beam-induced blue-shift observed for the isolated oscillator. To estimate the charge on the conductive cantilevers and considering that free electrons on the cantilever are repulsed by the electron beam, we assume that the charges accumulate at the cantilever edge that faces away from the electron beam, i.e. $R_o = d + w$, where $w$ is the cantilever width. The effective mass $m_{\text{eff}}$ is calculated from the geometrical and material parameters of the cantilever, which is about 5.7 pg. The spring constant $k$ was evaluated for both cases at the electron beam position that least affects the resonance frequency ($d = 300\ nm$), which were, $k = (2\pi f_o)^2 m_{\text{eff}} = 0.0249$ N/m for isolated cantilever and 0.0373 N/m for grounded cantilever. For 10 kV acceleration voltage, $v_e$ is about 5.93 x 10$^7$ m/s and the beam current $I_e$ is 1.5 nA. Fig. 4b shows the calculated charge as a function of the distance of the electron beam from the cantilever edge. Notably, it indicates that some charging also occurs in the grounded case (green line), which may be expected as some charge will accumulate in the cantilever's non-conductive layers.

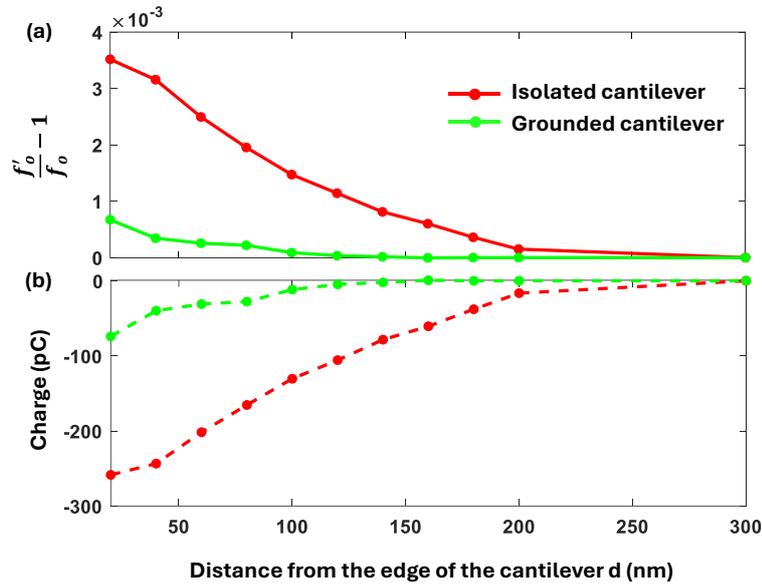

Figure 4: a) Relative frequency shift as a function of distance $d$ between the electron beam and the cantilever edge for both grounded and isolated cantilevers. (b) Charge implied by the observed frequency shift according to Eq. (11).

The charge $q$ estimated when the electron beam is 20 nm away from the edge of the isolated cantilever is $-258$ pC, which corresponds to 1.6 x 10$^9$ electrons and may be compared to the 7.5 x 10$^{10}$ gold atoms on the cantilever, implying that the charge corresponds to about 1 electron for every 47 gold atoms.

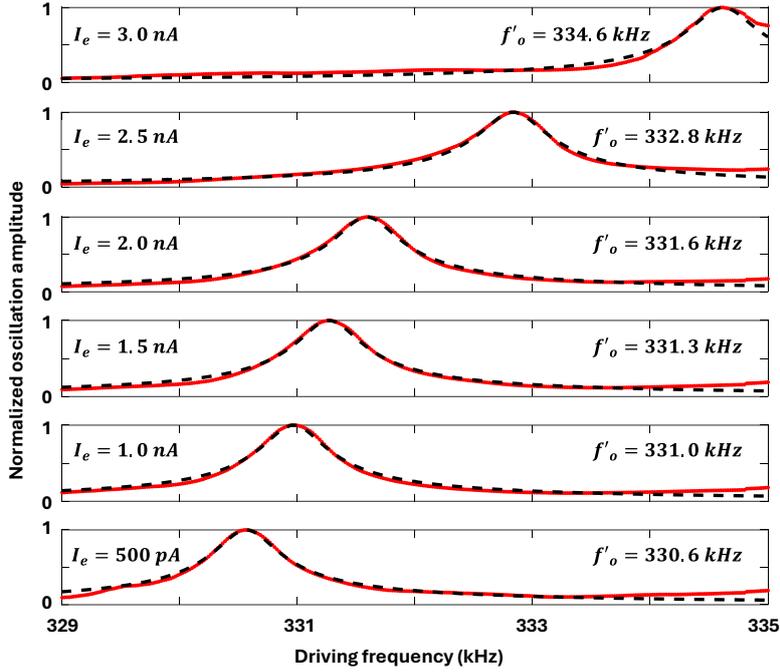

*Figure 5: Oscillation amplitude as a function of driving frequency for the isolated cantilever for different currents ($I_e$) of an electron beam placed at a distance of $d = 100\ nm$ from the cantilever edge. The cantilever's mechanical resonance blue-shifts with increasing beam currant.*

**The effect of electron beam current on a cantilever's resonance**

The analytical expression for modified resonance frequency (Eq. (8)) depends on experimental parameters that can be controlled, such as electron beam acceleration voltage that controls $v_e$, beam current ($I_e$) and the electron beam position. This gives us control over the resonance frequency of the isolated cantilever. Here, we fix the acceleration voltage to 10 kV and the electron beam injection position at $d$ = 100 nm. Under these conditions, the beam current was varied from 500 pA to 3 nA in 500 pA steps. We observe that the resonance frequency increases superlinearly with the beam current (Fig. 5). Given that the charge accumulated on the cantilever may be expected to be proportional to the electron beam current ($q \sim I_e$), we should expect a quadratic dependence of the resonance frequency shift on the electron beam current according to the approximate form of Eq. (8). If we also consider that an increased size of the electron beam for larger currents further contributes to charging, we may expect the resonance to shift faster than quadratically with electron beam current, and indeed the observed resonance shift is best fitted by $\Delta f_o \sim I_e^{3.4}$.

The dependence of mechanical resonance frequencies on the charge of the resonator and the presence of nearby charges or charged particle beams provides an opportunity for sensing. For example, resonance shifts of a mechanical resonator with a known charge may be measured to detect nearby charges or the charge of a particle beam. Or, the charge of a mechanical resonator may be determined from resonance shifts in the presence of a known nearby charged particle beam, as illustrated above.

**Methods**

**Sample fabrication.** A silicon nitride membrane of 50 nm thickness supported by a 200-μm-thick silicon frame (NORCADA), was coated with 50 nm of gold using a BOC Edwards resistance evaporator and then 50 nm of gallium nitride (GaN) by sputtering using an AJA Orion sputtering machine. (GaN is a cathodoluminescent material that was used to enable further experiments that are beyond the scope of this manuscript.) The nanomechanical cantilevers were fabricated by focused ion beam milling using a Helios Nanolab 600 SEM-FIB dual beam system. The cantilevers were designed using Design CAD Express 16, and milling was controlled using

NPGS software. The magnification was fixed at 2500x, centre-to-centre distance, and line spacing was set as 10 nm. The beam current used was 52.9 pC, and the area dosage was 50 mC/cm$^2$. Then, one of the otherwise identical cantilevers was isolated by FIB milling through the electrically conductive gold layer around its base, which is supported by the membrane's silicon frame, while the other cantilever remained grounded (Fig. 2a).

**Experimental characterization.** The experiment is conducted using a CamScan3600 SEM with a field emission gun. The sample was mounted inside the SEM chamber (3 μbar pressure) on a shear piezoelectric stack (Thorlabs PL5FBP3). The driving voltage signal to the piezo-actuator is provided from the signal output of a Zurich Instruments UHFLI 600 MHz lock-in amplifier (frequency sweep with sinusoidal waveform and 500 mV$_{pp}$ amplitude). The secondary electron signal is collected using an Everhart-Thornley secondary electron detector and detected using a signal input channel of the Zurich Instruments UHFLI 600 MHz Lock-in amplifier.

**Conclusion**

In summary, our work shows experimentally how the in-plane resonance frequency of a charged nanomechanical resonator depends on its distance from and the current of a nearby electron beam. A blue-shift of the mechanical resonance was observed and explained by an analytical model as a consequence of how the Coulomb electrostatic force modifies the effective spring constant of the resonator. The model fits the experimental data and provides an analytical estimate of the charge. The effect may be exploited for sensing of charges and control of mechanical resonances in nanomechanical systems as well as detection of the charge of charged particle beams.

**Acknowledgement.** This work was supported by the Engineering and Physical Sciences Research Council, UK (grant number EP/T02643X/1).

**Conflict of interest.** The authors state no conflict of interest.